\documentclass[aps,prb,twocolumn,groupedaddress,showpacs]{revtex4}
\usepackage{amsmath}
\usepackage{graphicx}
\usepackage{graphics}
\usepackage{color}
\usepackage{bbold}
\begin{document}
\title{Dynamics of domain-wall Dirac fermions on a topological insulator: a chiral fermion beam splitter}
\author{Ren\'{e} Hammer}\author{Walter P\"{o}tz}
\affiliation{Institut f\"{u}r Physik, Karl-Franzens-Universit\"{a}t Graz, Universit\"{a}tsplatz 5, 8010 Graz, Austria}

\date{\today}

\begin{abstract}
The intersection of two ferromagnetic domain walls placed on the surface of topological insulators provides a one-way beam splitter for domain-wall Dirac fermions.  Based on an analytic expression for a static two-soliton magnetic texture we perform a systematic numerical study of the propagation of Dirac wave packets along  such intersections. A single-cone staggered-grid finite difference lattice scheme is employed in the numerical analysis.  It is shown that the angle of intersection plays a decisive role in determining the splitting ratio of the fermion beam. For a non-rectangular intersection,  the width and, to a lesser extent,  the type of domain walls, e.g. Bloch or N{\'e}el, determine the properties of the splitter. As the ratio between domain-wall width and transverse localization length of the Dirac fermion is increased its propagation  behavior changes from quantum-mechanical (wave-like) to classical ballistic (particle-like).     An electric gate placed near the intersection offers a dynamic external control knob for adjusting the splitting ratio.  
\end{abstract}


\maketitle

\section{Introduction}\label{intro}

There has been considerable progress in the investigation of Dirac fermions on (textured) TI surfaces in external electromagnetic fields.   Experimentally, the Dirac cone spectrum with and without external magnetic field has been verified in recent experiments.\cite{analytis,chen1,xia}    Position control of the Fermi energy within the bulk gap of the insulator has become feasible.\cite{chen2,checkelsky,xiu,kim,wang,ren}   Magnetic texturing of TI surfaces has been explored experimentally.\cite{chen1,honolka,wei}  In one study using iron as the dopant, however,  the easy axis has been reported to be in plane.\cite{honolka}  In thin films of MBE grown Cr doped ternary (Bi$_x$Sb$_{1-x}$)$_2$Te$_3$ ferromagnetic order has been reported.\cite{chang}    
With the theoretically predicted existence of topologically protected surface states confirmed in experiment, one of the next steps is to learn to manipulate them by structural design and to utilize their properties in chiral fermion devices.   On the theoretical side, numerous papers have been published on related topics, such as magneto-transport on TI surfaces in presence of  ferromagnetic layers, \cite{yokoyama,soodchomshom,yuan,kong,suwanvar,wu}  spin-polarized magnetic impurities on TI surfaces and Landau levels,\cite{misawa,vazifeh,schwab,nomura,burkov} the interaction of magnetic domain walls with Dirac fermions,\cite{loss,chui,castro}  lensing effects and transport perpendicular to domain walls,\cite{hassler,wickles} crossover from weak anti-localization to weak localization,\cite{liu2} and gate control of TI channel states.\cite{yokoyama2}  An interferometer for chiral fermions has been proposed recently.\cite{hammerAPL}

Surfaces of 3D topological insulators can be interpreted as a 2D domain wall between a spin-orbit driven band inversion within the insulator and normal band ordering in vacuum.\cite{qi,fu,moore,zhang}
The topologically protected gapless surface states in form of helical-state Dirac cones can be manipulated further by time-reversal-symmetry-breaking (TRB) perturbations applied to the surface.\cite{qi,hasan,liu}
Indeed, it has been known for quite a while that an effective-mass inversion domain wall can produce a 1D chiral edge eigenstate for a 2D Dirac fermion Hamiltonian.\cite{jackiw}  Such edge states are required to observe a quantized Hall effect.  
TRB can be induced by exposure to an (effective) external magnetic field.  For example, a ferromagnetic tip or permanent magnetic texture arising from the proximity of ferromagnets, in conjunction with magnetic doping of the surface,  may be used to induce sign changes in the mass term of the effective Dirac equation.\cite{zhang,honolka,wei,liu,chen1}  The expected order-of-magnitude for the mass-gap in materials, such as Bi$_2$Se$_3$, is up to several tens of meV.\cite{chen1,luo}

Nanostructuring of TIs enhances the surface over the bulk contributions to fermion charge transport and provides another promising means to produce edge states, following strategies previously applied to graphene.  It should be recalled that the properties of the surface states are a consequence of the TI bulk properties.  Bulk doping can be used to manipulate surface state behavior.\cite{wray:2011,xiu:2011,beidenkopf:2011,zhang:2012}  Compositional tuning of Dirac fermion electronic structure has been demonstrated for  BiTl(S$_{1Ðx}$Se$_{d}$)$_2$,  Bi$_2$(Te$_{3-x}$Se$_x$). \cite{xu-2011,chen:2013}  Placed on a substrate, electric contacts have been made to the TI.\cite{xiu:2011,chang:2013}   Metal-TI junctions have been studied theoretically.\cite{modak:2012}  Such structuring can be expected to lead to the realization of interesting quantum interference effects due to the helical nature of surface states, implying spin-polarized electric currents, the absence of back scattering, and robustness to moderate disorder.    This strategy has lead to the experimental observation of the quantum anomalous Hall effect in a magnetic topological insulator.\cite{chang:2013}   Quantum oscillations in TI nanoribbons in conjunction with high surface conductance (e.g., for Bi$_2$Te$_3$) have been investigated theoretically and in experiment.\cite{peng:2010,ihn:2010,xiu:2011}

In this paper  the dynamic properties of chiral fermions in channel states introduced by magnetic texturing of the surface of a TI insulator are explored.  We perform a theoretical study of the propagation of chiral domain-wall fermions along the intersection of two ferro-magnetic domain walls imprinted upon the surface of a TI.  In a systematic numerical analysis,  we extract the one-way beam splitting properties regarding magnetic texturing, such as the angle of intersection, domain-wall thickness, and details of in-plane magnetization.  
Our time-dependent analysis is based on a newly developed scheme for a numerical treatment of the (2+1)D Dirac equation in presence of electromagnetic fields, whose numerical mathematical properties will be presented in a forthcoming publication.  Here, the analysis is performed using static solitonic domain-wall crossings, valid in the adiabatic regime.  The paper is organized as follows.  In Sect. \ref{CHIRAL}  we give a brief summary of chiral (domain wall) fermions in external electromagnetic fields, as relevant for TI surfaces and introduce 2-soliton magnetic textures.  In Sect. \ref{MOD-H} we give the model Hamiltonian for our numerical analysis.  Sect. \ref{SCONE} gives a summary of our single-cone numerical lattice model for (2+1) Dirac fermions.  Sect. \ref{NUMR} features some of our  numerical results  for chiral fermion beam splitters.  Finally, a general summary and our conclusions are given in Sect. \ref{SUC}.

\section{Chiral domain-wall fermions}\label{CHIRAL}

\subsection{Basic considerations}

An effective (2+1)D model for the dynamics of Dirac fermions on a magnetically textured TI surface may be founded upon a Hamiltonian
\begin{equation}H=H_F+H_I+H_{FI}~,\end{equation}
 consisting of the fermion Hamiltonian $H_F$, the impurity Hamiltonian $H_I$,  and the interaction $H_{FI}$. 
For, the Dirac fermions in an external electromagnetic field one may write
\begin{equation}
H_F =  v\left( \boldsymbol\sigma\times  {\bf \Pi}  \right)\cdot \mathbf{\hat z} + \mu_B g_F \boldsymbol\sigma {\bf B}(x,y,t) + V(x,y,t)
\label{DH}
\end{equation}
Here, ${\bf \Pi}= {\bf p} +\frac{e}{c} {\bf A}(x,y,t)$ in the spin-orbit term denotes the kinetic momentum in presence of a vector potential ${\bf A}$ associated with the external magnetic field ${\bf B}(x,y,t)$,  $V(x,y,t)=-e\Phi(x,y,t)$ is the scalar potential energy, and 
$\mu_B=\frac{e\hbar}{2mc}$ and $g_F$ in the Pauli term are, respectively, the electron Bohr magneton and the Land{\'e} factor.   
 The Pauli vector  is proportional to the physical fermion spin with its direction-locked perpendicular to the particle current density and the normal vector to the surface $\mathbf{\hat z}$.\cite{qi}
Note, that the simple form of the spin-orbit term can be extended to more precisely represent the energy dispersion away from the Dirac point, such as an account of hexagonal warping.\cite{qi,alpichshev}  

The impurity spins may be modeled by a generic Heisenberg-type Hamiltonian of the form
\begin{equation}
H_I=-\sum_{i,j} J_{ij} {\bf S}_i {\bf S}_j - \sum_i g_I\mu_B {\bf S}_i {\bf B}^o_i ~, 
\end{equation}
and the Dirac fermion-impurity interaction by
\begin{equation}
H_{FI}=-\sum_i J'_i {\bf S}_i  \boldsymbol\sigma ~.
\end{equation}
Here ${\bf S}_i $ and ${\bf B}^o_i$ denote, respectively, the impurity spin and external magnetic field, for impurity site $i$.  Note that one must differentiate between the "external" magnetic field experienced by fermions and impurities.

The physical situation envisioned and captured by the Hamiltonian $H$ is that of a TI surface which is densely covered by magnetic impurities which interact with one another, as well as with the Dirac fermions under an exchange interaction.   Several possible origins for an exchange interaction $J_{ij}$ have been discussed in the literature.\cite{rosenberg,luo}  An external magnetic field may be applied to imprint and stabilize domain-wall formation between ferromagnetic ordered domains.  

Subjecting $H_I+H_{FI}$ to a mean-field approximation one obtains $H_I+H_{FI} \rightarrow \left(H_I+H_{FI}\right)^{MF} = -\sum_i g_I\mu_B {\bf B^{(I)}}_i$, where the effective magnetic field at impurity site $i$ is given as
\begin{equation}
 {\bf B}^{(I)}_i = {\bf B}^o_i +\frac{1}{g_I \mu_B}  \left[\sum_j J_{ij} \langle {\bf S}_j\rangle + \langle J'_i \boldsymbol\sigma \rangle \right]  ~.
\end{equation}
The three contributions arise from the external magnetic field (including contributions from the orbital motion of Dirac fermions), impurity magnetization, and fermion spin polarization.  The latter gives rise to a spin-transfer torque.

Similarly for the fermions,  
\begin{eqnarray}
H_F  & + &H_{FI} \rightarrow  \left(H_F+H_{FI}\right)^{MF} \label{DHMF} \\
& =  & v\left( \boldsymbol\sigma\times  {\bf \Pi}'  \right)\cdot \mathbf{\hat z} + \mu_B g_F \boldsymbol\sigma {\bf B^{(F)}}(x,y,t) + V(x,y,t)~, \nonumber  \label{HFMF}
\end{eqnarray}
where the effective magnetic field in the Pauli term is given by 
\begin{equation}
{\bf B}^F(x,y,t)= {\bf B'}(x,y,t) - \frac{1}{g_F \mu_B}\sum_i \langle J'_i {\bf S}_i \rangle \boldsymbol\sigma ~, \label{BFE}
\end{equation}
and the vector potential ${\bf A}'$ entering  the canonical momentum ${\bf \Pi}'$ in the spin-orbit term contains the contribution from the external magnetic field and the magnetization of the impurities, as will be discussed below, such that  ${\bf B'}(x,y,t)=\boldsymbol\nabla \times {\bf A}'(x,y,z,t)\mid_{z=0}$.
 
In what follows we concentrate on the Dirac fermion dynamics.  The impurity dynamics, in principle, can be treated self-consistently in parallel.  However, it is generally accepted that the latter occurs on a time-scale which is long compared to the fermion dynamics and adiabatic schemes have been used successfully to model the interplay between the two subsystems.\cite{wenin} 

 According to Eq. \eqref{DHMF}  the presence of magnetic impurities and a stabilizing external $B$-field has two consequences for the  dynamics of Dirac fermions:  an impurity spin polarization (magnetization)  {\bf M} modifies the net external magnetic field from ${\bf B}$ to ${\bf B}'$ and, for $M_z\neq 0$,  introduces an exchange term  ("mass term"), in addition to the Zeeman term, in Eqs. \eqref{DHMF} and \eqref{DH} below.
Such a mass term has been estimated to be of the order of up to several tens of meV and represents the dominant magnetic-field contribution.\cite{luo}  For a fermion g-factor $~20$ the effective magnetic exchange field required for a mass gap of $~25$ meV is about 10 T.\cite{liu,lande} 
Such a magnetic exchange field will, as usual for ferromagnets, dominate  any typical static external field ${\bf B}\leq 0.5$ T in Eq. \eqref{BFE} in the Pauli term of Eq. \eqref{DHMF}.  

In order to estimate the magnetization {\bf M} associated with an exchange field of 10 T, we use parameters typical for Mn impurities.\cite{vanesch}  Within the simple form $J'_i=J'\delta(x-x_i)\delta(y-y_i)$ 
and $J'\approx 130$ meV nm$^2$ one needs an impurity density of about 0.1 nm$^{-2}$ to achieve a mass gap of $~25$ meV.  Using $g_I\approx 2$, the magnetization per area is of the order of $1.2\times 10^{-6}$ meV/(Gauss nm$^2$). 
Based on this estimate and a layer thickness of 1 nm, the order of magnitude of the magnetization contribution to $B'$, $|B'-B|=4\pi M$,  is about $2.5\times 10^{-3}$ T, making this effect negligible in both the spin-orbit and Pauli term of Eq. \eqref{DHMF}. 
In principle there also is a topological field contribution due to the magneto-electric effect, but it is extremely small in magnitude ($\leq 10^{-6}T$).\cite{qi, garate}
In summary, the dominant magnetic-field effect onto the Dirac fermions arises from the exchange field, followed in importance by the external magnetic field in the spin-orbit term.  
The effect of an external magnetic field on the spectrum of TI Dirac fermions has been investigated both theoretically and experimentally (see Introduction \ref{intro}).
It  has a negligible effect on the dynamics of domain-wall Dirac fermions when compared to the in-plane component of the exchange term, as we have verified numerically in our studies detailed below.   We conclude that the formation of domain-wall states and the dynamics of domain-wall Dirac fermions is dictated predominantly by the magnetic domain-wall structure (exchange field) and the external electric bias.   
However, the presence of an external magnetic field may be essential to pin domain walls and to imprint and stabilize a specific ferromagnetic domain-wall structure.

\subsection{Solitonic magnetic textures}

The rich physics of ferromagnetic domain wall dynamics has been well documented in the literature.\cite{how,yokoyama:2010,loss,chui,castro}  
Here we consider well-pinned hard ferromagnetic textures and explore domain-wall fermion dynamics on  a time-scale which allows for a quasi-static treatment of the domain wall structure, neglecting spin-transfer torque effects.\cite{wenin}  The latter implies the regime of low current density.   Our numerical approach can handle time-dependent magnetic domains, however, such an analysis will be the topic of future investigations.

An analytic model for stable, local-minima, 2D magnetic textures can be derived  
from solutions to a simple free energy functional which consists of an isotropic exchange term and an anisotropy term with $z$ as the easy axis.  In terms of the unit magnetization ${\bf m} = (m_x,m_y,m_z)=(\sin\theta \cos\phi,\sin\theta \sin\phi, \cos\theta)$  and with  $a,b>0$ it is
\begin{equation}
{\cal F} ({\bf m})= \frac{1}{2} \int dx dy \left[ a (\mbox{\bf grad} {~\bf m})^2 - b m_z^2\right]~. \label{Fmag}
\end{equation}
In spherical coordinates $\theta(x,y)$ and $\phi(x,y)$  ${\cal F}$ takes the simple form 
\begin{eqnarray}
{\cal F} ({\bf m}) & = & \frac{1}{2} \int dx dy [ a \sum_{i=x,y} \left((\partial_{x_i} \theta)^2 + (\partial_{x_i} \phi \sin{\theta})^2\right)  \nonumber \\
& &  - b    \cos(\theta)^2]~.
\end{eqnarray}

Optimality conditions lead to $\partial_{x_i} \phi =0$ and 
\begin{equation}
\theta_{xx} +  \theta_{yy} =\frac{b}{2a}\sin{2 \theta}.
\end{equation}
Setting $u=2\theta$ and rescaling 
$x_i \rightarrow \sqrt{b/a} x_i $ one obtains the sine-Gordon  equation (in "imaginary time" $i y$)
$$
 u_{xx} + u_{yy}  =\sin{2 u}.
$$
Multi-soliton solutions are most directly found using Hirota's bilinear expansion in $\epsilon$.\cite{hirota,bowtell,zarkharov}
The single-soliton solution takes the form
\begin{eqnarray}
\theta^{(1)}(x,y) & = & 2\arctan\left\{ e^{ \eta(x,y)}\right\},  \nonumber \\
\eta(x,y)& = & \sqrt{b/a}(x'-x_o'),
~ x_o'=-\sqrt{a/b}\ln \epsilon   \nonumber \\
 x' & = & x \cos{\beta}  + y \sin{\beta} , ~ \beta \in [0,2\pi] \nonumber \\
 \phi(x,y) & = & \phi_o, ~\phi_o  \in [0,2\pi] \label{1S} .
 \end{eqnarray}
This solution represents a magnetization domain wall along $y'=- x \sin{\beta} + y  \cos{\beta}$ at  $x_o$.  

A magnetic domain-wall intersection which represents a local minimum to the free energy functional Eq. \eqref{Fmag} can be constructed from the two-soliton solution
\begin{eqnarray}
\theta^{(2)}(x,y) & = & 2\arctan\left\{ \frac{e^{ \eta_1(x,y)}+e^{ \eta_2(x,y)}}{1+\kappa_{12} e^{ \eta_1(x,y)+\eta_2(x,y)}}\right\},  \nonumber \\
\eta_i(x,y)& = & \sqrt{b/a}(x_i-x_o^i),
 \nonumber \\
 x_i & = & x \cos{\alpha_i}  + y \sin{\alpha_i} , ~ \alpha_i \in [0,2\pi], i=1,2 \nonumber \\
 \kappa_{12}& = & \frac{(\cos{\alpha_1}-\cos{\alpha_2})^2+(\sin{\alpha_1}-\sin{\alpha_2})^2}{(\cos{\alpha_1}+\cos{\alpha_2})^2+(\sin{\alpha_1}+\sin{\alpha_2})^2} \nonumber \\
 \phi(x,y) & = & \phi_o, ~\phi_o  \in [0,2\pi]~.\label{2S}
 \end{eqnarray}
 We place the point of intersection at the coordinate origin, setting $x_o^1=x_o^2=0$, retaining $\alpha_1-\alpha_2(\neq 0,\pi)$ and $\phi_o$ to characterize the type of intersection.
With suitable magnetic impurities present at or near the surface of the TI, such a ferromagnetic domain wall intersection may be induced by an array of magnetic poles of the form $\left[\begin{array}{cc} + & -\\- &+\end{array}\right]$ facing the TI surface.

\begin{figure} [!t]
\includegraphics[width=8.6cm]{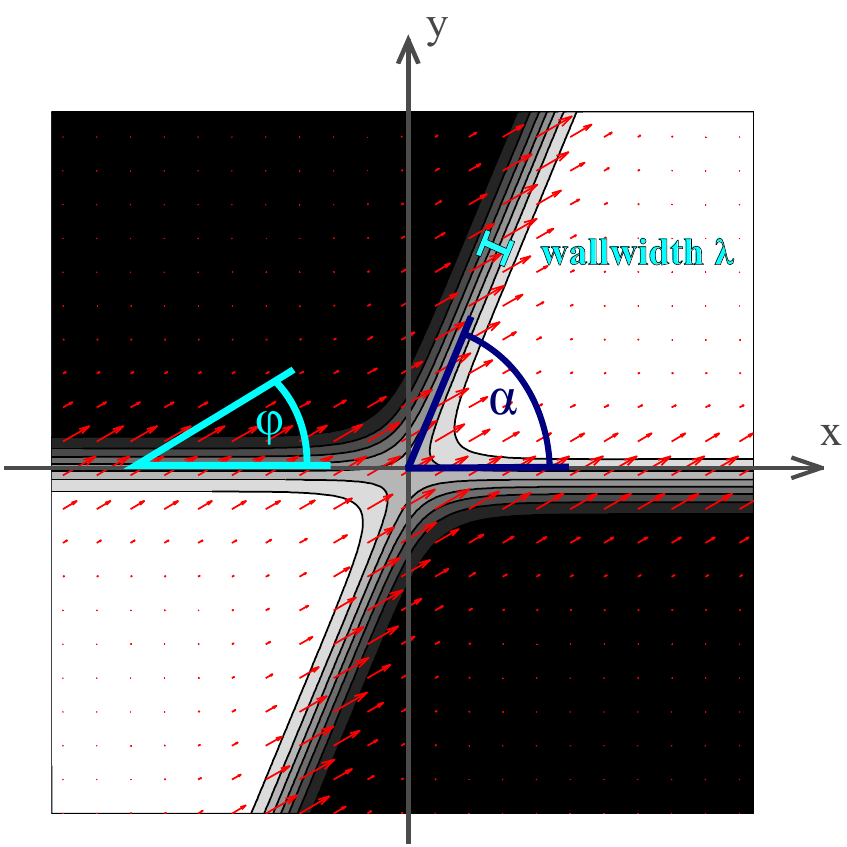}
\caption{(Color online). Magnetic texture of a domain wall intersection. The contour plot encodes the $M_z$ component of the magnetization,  asymptotically taking the values $\pm \mid {\bf M}\mid $  (black and white regions). The arrows show the in-plane components $(M_x,M_y)$. $\alpha$ is the angle of intersection and $\phi$ is the angle of the in-plane magnetization direction relative to the x-axis. The wall width is given by $\lambda = \sqrt{a/b}$, where $a$ is the exchange constant and $b$ the anisotropy parameter (see main text).}
\label{texture}
\end{figure}

\subsection{The model Hamiltonian and domain-wall states}\label{MOD-H} 

According to Sect. \ref{CHIRAL} the effective Hamiltonian for a single Dirac cone on the surface of the TI interacting with magnetic texture $\mathbf{M}(x,y)$ may be written\cite{qi,zhang,shan} 
\begin{equation}
H =  v\left[\boldsymbol\sigma\times {\bf \Pi}(x,y,t)\right] \cdot \mathbf{\hat z}+ \mathbf{M}(x,y)\cdot\boldsymbol\sigma + V(x,y)~.
\label{DH1}
\end{equation}
According to Eq. \eqref{HFMF} second and third term on the r.h.s., respectively, account for the presence of the magnetic texturing 
${\bf M}(x,y) = \hat{{\bf J}}  {\bf m}(x,y) $, with $ \hat{{\bf J}}$ denoting the effective exchange coupling tensor,  and an external scalar electric potential arising, for example,  from a gate bias.  

This effective Hamiltonian describes the 2D excitation spectrum near the Dirac point, including 1D edge states in presence of domain walls.  Domain-wall edge states are responsible for an integer quantum Hall effect on the TI surface, as discussed in the literature.\cite{qi,zhang}   For $H$ Eq. \eqref{DH1} and  a 
single-soliton domain wall Eq. \eqref{1S}, with 
$\lim_{x'\rightarrow \pm \infty} M_z(x') =M_{\pm}$ and  $M_{-} M_{+}<0$,  a domain-wall eigenstate exists which takes the 
simple form\cite{hasan} 

%
\begin{equation}
 \langle x,y \left|Y' ,\pm \right\rangle \propto  \frac{1}{\sqrt{2}} \begin{pmatrix}   \pm e^{- i\alpha/2} \\  e^{i\alpha/2}  \end{pmatrix}e^{\mp \frac{1}{\hbar v} \int_{x'_o}^{x'} M_z(x'') dx'' +ik_{y}' y'}\;,
 \label{Y'}
\end{equation}
when neglecting in-plane contributions to the magnetization (valid when sufficiently far away from the domain wall).\cite{hammerAPL}
It features a linear dispersion  $E=\pm v \hbar\; k_{y}'$, with the upper sign for  $M_- < 0$ and $M_+ > 0$, and the 
 lower sign for $M_+ < 0$ and $M_- > 0$.  Note that a constant potential $V$ in Eq.\eqref{DH} simply adds to the eigenvalue
 $E$.    Parallel and curved zero-mass lines have also been studied in the literature.\cite{tudorovskiy:2012}  For a general form of an in-plane magnetization  $M_x(x'), M_y(x') \neq 0$  or other more complicated domain-wall structures eigenfunctions are best found numerically.  
 
More complex magnetic texturing, where islands of positive mass neighbor islands of negative mass, produces a network of one-way chiral channel states.\cite{hammerAPL}  Here we study an elementary building block of such a network in form of an intersection of two linear domain walls.
With the realization of such a texture a reflection-less beam splitter for chiral fermions is established.  This is demonstrated numerically below for the two-soliton texture  Eq. \eqref{2S}.
\\

\section{Numerical method: a single-cone lattice model}\label{SCONE}

\begin{figure}[h!]
\centering
\includegraphics[width=7cm]{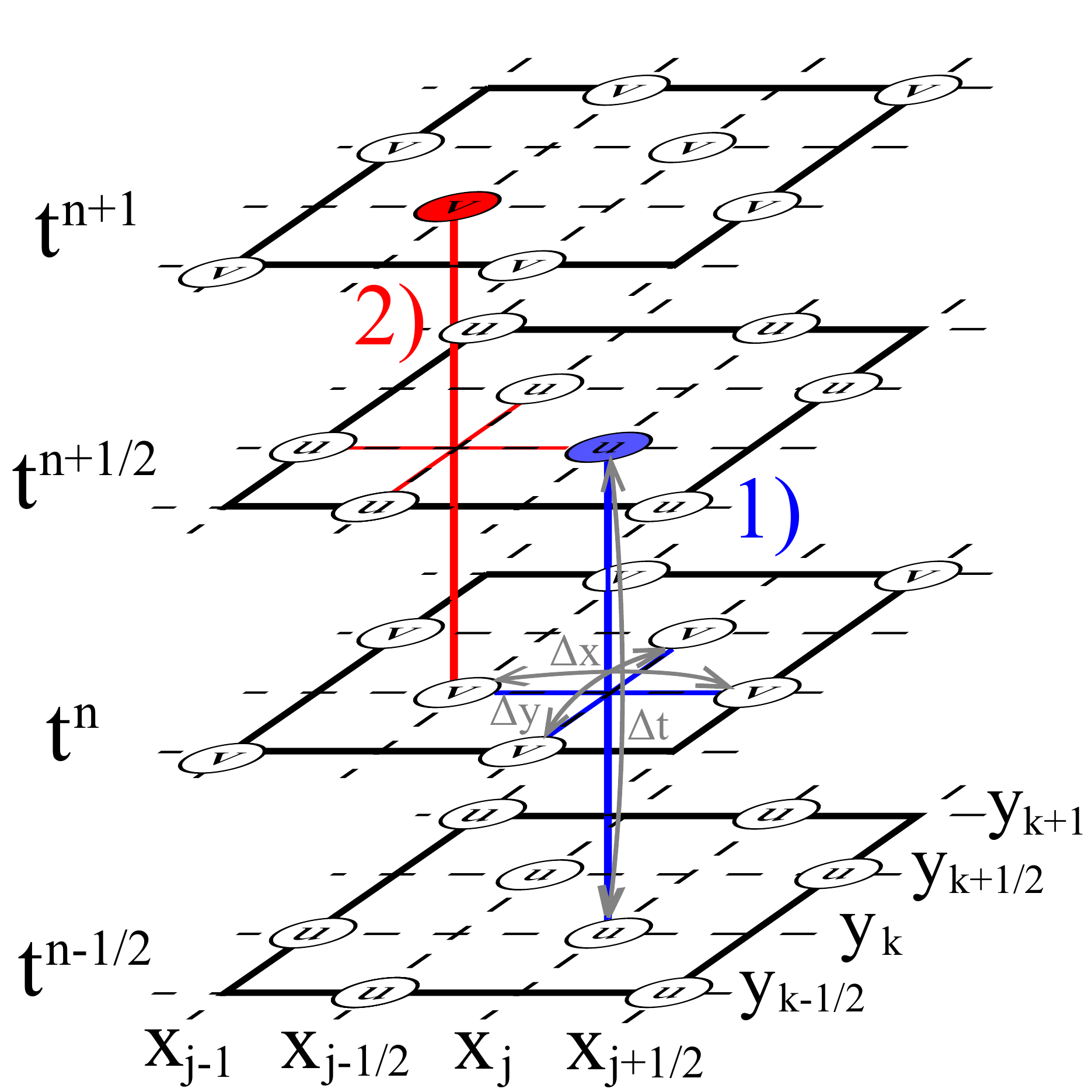}
\caption{Leap-frog time stepping on a time and space staggered grid for the (2+1)D Dirac equation.}\label{scheme-fig}
\end{figure}

The time-dependent Dirac equation is solved numerically  for $H$ in Eq. \eqref{DH}.   
Putting the (single cone) Dirac equation onto a grid for numerical solution traditionally has been hampered by fermion doubling, that is,  the lattice model has more eigenmodes than the original continuum model.  Two different numerical finite difference schemes have been employed and compared in the course of this analysis.  Both use a staggering of the spinor components in space and time. The first one has the following advantageous features:\cite{submCPC}  (i) it provides the exact (linear) dispersion relation for mass-less free Dirac fermions  along the main axes $k_x$ and $k_y$, (ii) it allows an implementation of absorbing boundary conditions via an imaginary potential term, and (iii) it allows for a removal of the second Dirac cone, located at the corners of the Brillouin zone,  by a Wilson term.\cite{wilson}  The second one, briefly outlined below, avoids the fermion doubling problem altogether at the cost of loosing the 
perfect (i.e., linear) dispersion property along the main coordinate axes for mass zero.\cite{nielsen} It features a single Dirac cone dispersion without the need for using a Wilson mass term to get rid of the doublers. For simple rectangular magnetic structures aligned with the grid's $x$ and $y$ axis the first scheme has higher accuracy, for general setups, however,  the second scheme performs better. The figures shown below are obtained with the second scheme. Results from various simulations within the first scheme have and will be presented elsewhere.\cite{hammerAPL,arxiv,submCPC} Global grid refinement experiments where done and then the simulations where executed with a grid for which a further halving of the grid-spacings gave an improvement no more than $1\%$. For comparison some simulations were done with scheme one showing a difference in the result for the transmission of less than $1\%$.\\
\\ 
The finite difference scheme for the Dirac equation $i \partial_t \mbox{\boldmath$\psi$}(x,y,t) = \hat{H} \mbox{\boldmath$\psi$}(x,y,t)$ with the Hamiltonian Eq. \eqref{DH}, where $\mbox{\boldmath$\psi$}(x,y,t)\in\mathbb{C}^2$ is a $2$-component spinor, may be summarized as follows.   Introducing the space-time staggered according to in Fig. \ref{scheme-fig}  for the components of the spinor $\mbox{\boldmath$\psi$} = (u,v)$ and using symmetric second order accurate approximations for the derivatives we propose the following discretization of the (2+1)D Dirac equation
\\
\begin{align}
&\frac{u_{j,k}^{n+1/2}-u_{j,k}^{n-1/2}}{\Delta t}=\nonumber\\
&-i\big[(M_z)^n_{j,k}-V^n_{j,k}\big]\frac{u_{j,k}^{n+1/2}+u_{j,k}^{n-1/2}}{2}\nonumber\\
&-\frac{(v_{j+1/2,k}^{n}-v_{j-1/2,k}^{n})}{\Delta x}+i\frac{(v_{j,k+1/2}^{n}-v_{j,k-1/2}^{n})}{\Delta y}~,\nonumber\\
&-(M_x)^n_{j,k}\frac{(v_{j+1/2,k}^{n}+v_{j-1/2,k}^{n})}{2}\nonumber\\
&-i (M_y)^n_{j,k}\frac{(v_{j,k+1/2}^{n}+v_{j,k-1/2}^{n})}{2}~,\nonumber\\
\nonumber\\
&\frac{v_{j-1/2,k}^{n+1}-v_{j-1/2,k}^{n}}{\Delta t}=\nonumber\\
&+i\big[(M_z)^{n+1}_{j-1/2,k}+V^{n+1}_{j-1/2,k}\big]\frac{v_{j-1/2,k}^{n+1}+v_{j-1/2,k}^{n}}{2}\nonumber\\
&-\frac{(u_{j,k}^{n+1/2}-u_{j-1,k}^{n+1/2})}{\Delta x} -i \frac{(u_{j-1/2,k+1/2}^{n+1/2}-u_{j-1/2,k-1/2}^{n+1/2})}{\Delta y}\nonumber\\
&-(M_x)^{n+1}_{j-1/2,k}\frac{(u_{j,k}^{n+1/2}+u_{j-1,k}^{n+1/2})}{\Delta x} \nonumber\\
&+i (M_y)^{n+1}_{j-1/2,k}\frac{(u_{j-1/2,k+1/2}^{n+1/2}+u_{j-1/2,k-1/2}^{n+1/2})}{\Delta y}
~.\label{scheme}
\end{align}
The $u$-component defined for the discrete time indices $n-1/2\in \mathbb{Z}$ `lives' on the discrete space grid points $(j,k)\in\mathbb{Z}^2$ and $(j-1/2,k-1/2)\in\mathbb{Z}^2$, while the $v$-component defined for $n\in \mathbb{Z}$ is defined for space indices $(j-1/2,k)\in\mathbb{Z}^2$ and $(j,k-1/2)\in\mathbb{Z}^2$.\\
\\
\noindent The dispersion relation for constant coefficients is revealed using a plane-wave ansatz $u^{n+1}_{j+1,k+1} = e^{i (\omega \Delta t - k_x \Delta x - k_y \Delta y)}u^{n}_{j,k}$ (and analogously  for $v$). The centered approximation for the time and space derivatives, respectively,  translates into a multiplication by $\frac{2 i}{\Delta t} \sin \frac{\omega \Delta t}{2}$ and $\frac{2 i}{\Delta x,y} \sin \frac{k_{x,y} \Delta x,y}{2}$.  Time averaging leads to the factor $\cos \frac{\omega \Delta t}{2}$.
Solving for $\omega$ gives the dispersion relation for $M_x=M_y=V=0$
\begin{align}
\omega =& \pm \frac{2}{\Delta t} \arcsin \Bigg[ \frac{\Delta t}{2 +(M_z)^2 \Delta t}\nonumber\\
&\sqrt{(M_z)^2 + \Big(\frac{2}{\Delta x}\sin \frac{k_{x} \Delta x}{2}\Big)^2 + \Big(\frac{2}{\Delta y}\sin \frac{k_{y} \Delta y}{2}\Big)^2}\Bigg]~.\label{dispersion}
\end{align}

\noindent The dispersion relation for $M_x=M_y=V=0$ is monotonic and has its single minimum at $k_x=k_y=0$.
\begin{figure}[!t]
\includegraphics[width=8cm]{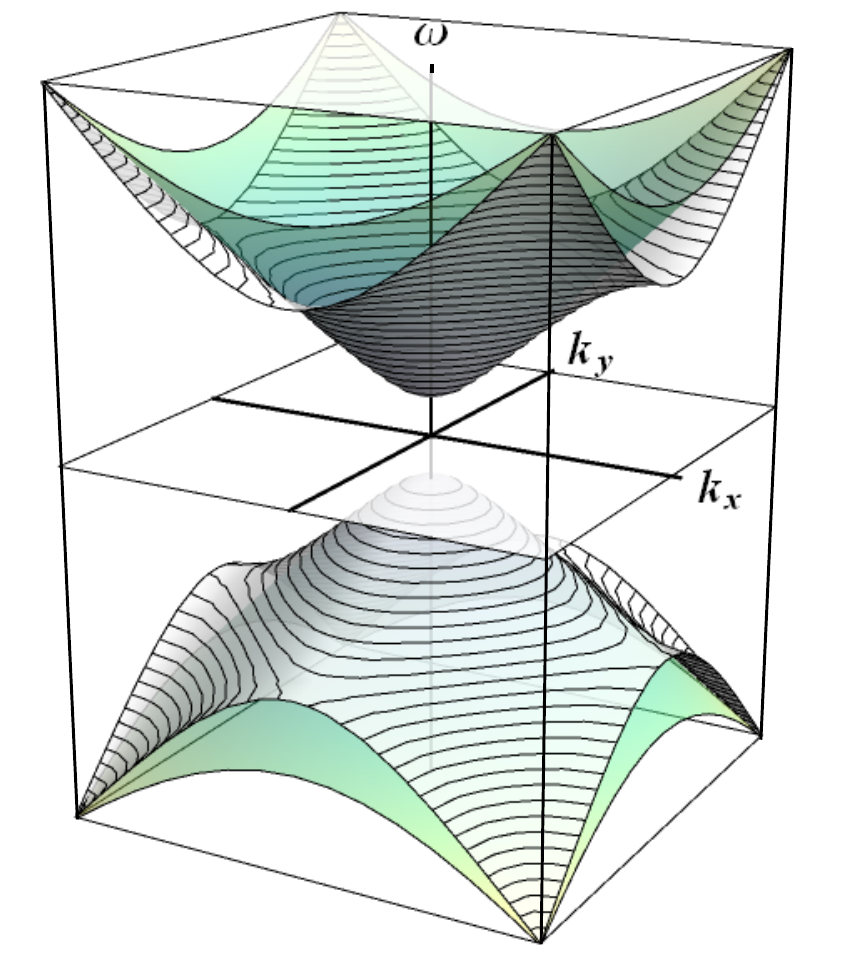}
\caption{(Color online). The dispersion within the numerical approach for a mass gap $2M_z = 1$ compared to the  exact cone of the continuum approach (in green). The grid spacings are chosen to be $\Delta_x=\Delta_y=1$ and $\Delta_t = 1/\sqrt{2}$.}\label{numdisp}
\end{figure} 
On the grid the $k$ vectors are defined up to reciprocal lattice vectors, only, 
leading to $k \in (-\pi / a, \pi/a]$, where $a = \Delta x, \Delta y$.
Accordingly the domain for the frequency is $\omega \in (-\pi / \Delta t, \pi/\Delta t]$.   Fig. \ref{numdisp} compares the dispersion within the (2+1)D lattice model for the lattice parameters $\Delta_x=\Delta_y=1$ and $\Delta_t = 1/\sqrt{2}$ and a mass gap $2m = 1$ to the exact cone of the continuum model (in green). 
A more detailed analysis of the scheme including a rigorous stability analysis (as performed recently for the (1+1)D case)\cite{1D} for general time- and space-dependent mass vector $\mathbf{M}$ and electro-magnetic potentials will be given elsewhere.\cite{Hammerup}

\begin{figure}[!t]
\includegraphics[width=8cm]{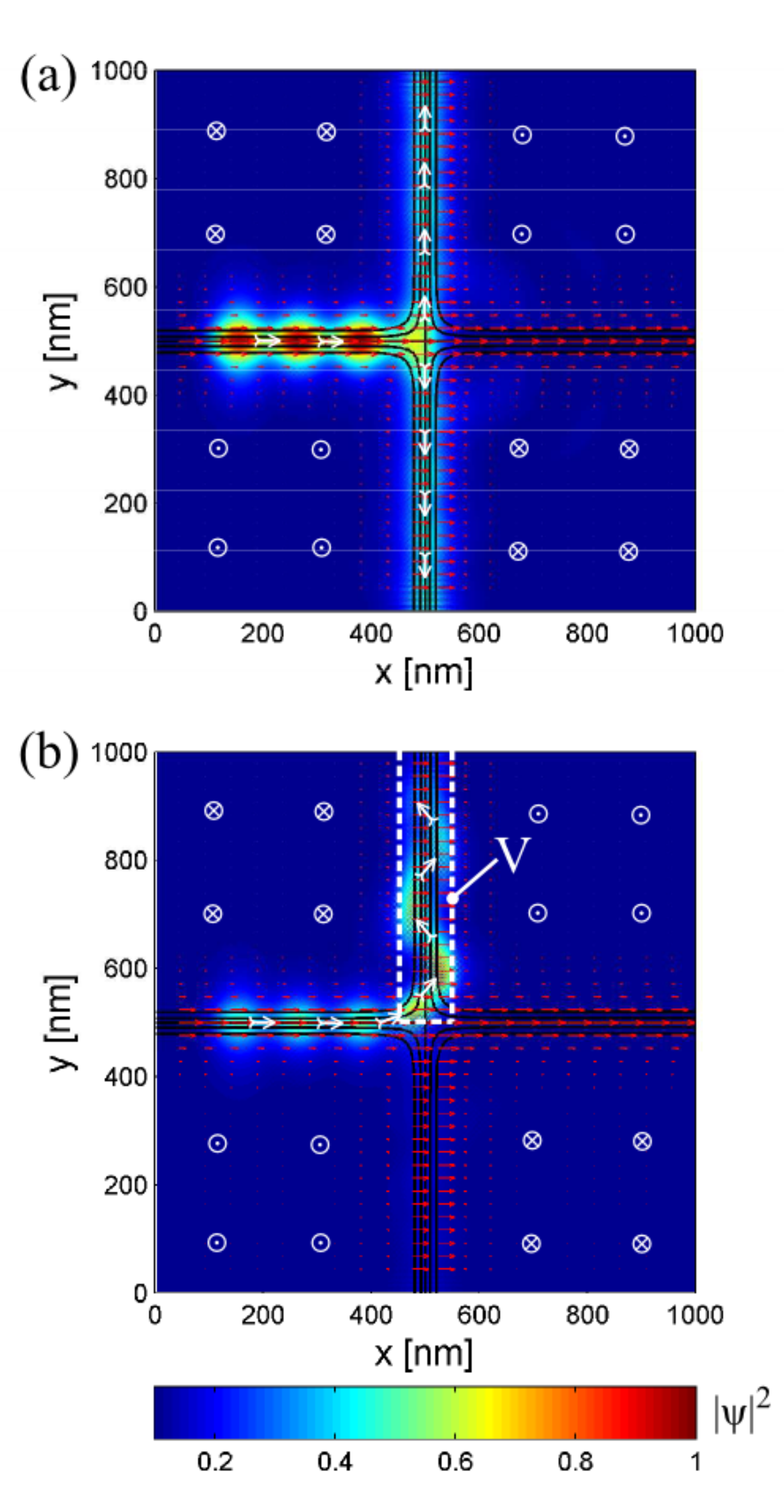}
\caption{(Color online). Snapshots of the wave packet propagation in a $90$-degree domain wall intersection: (a)  unbiased, (b) biased ($15$ mV).  The color (or brightness variation) shows the probability density $|\psi|^2$ (see color-bar). The black contour lines and the white crossed and dotted circles show the $M_z$-component of the magnetization. The vector plot shows the direction and magnitude of the in-plane magnetization $(M_x,M_y)$.}\label{wppwithV}
\end{figure} 
\begin{figure}[!t]
\includegraphics[width=8.6cm]{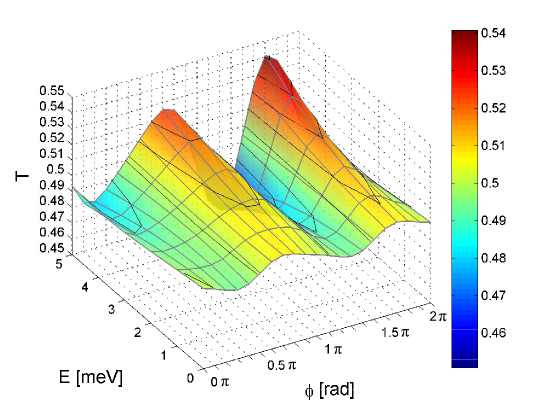}
\caption{(Color online). Relative transmission to the upper channel in a rectangular domain wall intersection with wall width $\lambda = 25$ nm as a function of the in-plane magnetization angle $\phi$ and the energy mean value $E$ of the  wave packet.}\label{is1}
\end{figure} 
\begin{figure} [t!]
\includegraphics[width=8.6cm]{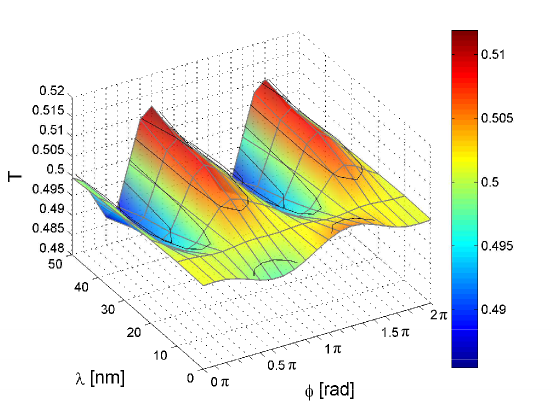}
\caption{(Color online). Relative transmission to the upper channel in a rectangular domain wall intersection as a function of the in-plane magnetization angle $\phi$ and wall width $\lambda$. The in propagation-direction Gaussian initial wave packet is prepared with an energy mean value of $E=0$.}\label{is2}
\end{figure}

\section{ A Dirac fermion beam splitter: numerical results}\label{NUMR}

The asymptotic value for the z-component of the magnetization in Eq. \eqref{DH}
 is chosen to be $\mid m_z\mid =7.5$ meV.  This value  is realistic in view of the expected exchange coupling.\cite{luo}
Note that the results obtained for this value can be scaled to other values due to the scale invariant nature of the problem at hand.  Indeed,  the relevant parameters characterizing a physical situation can be separated into ratios and scale-free (absolute) parameters. The relevant ratios are: the ratio of the confinement length of the wave packet to the domain wall width, the ratio of the wave packet energy to the gap (established by the asymptotic value of the z-component of the magnetization), and the ratio of the in-plane to the out-of-plane exchange coupling constant (in principle, the in plane component can also have an anisotropy). The scale free parameters are the angle of intersection and the angle $\phi$ for the in plane magnetization direction (see Fig. \ref{texture}).  For real structures in experiment sample-specific  imperfections, such as unwanted irregularities in the magnetic structure, may play a role. 

A typical simulation region of $1000 \times 1000$ nm is used and, for the simulations to follow,  we place an initial Gaussian  wave packet in the in-channel (see also  Fig. \ref{texture}).    It is characterized by its energy mean value $E$ and standard deviation, see  Fig. \ref{is1}.  Its initial shape perpendicular to the channel is given by Eq. \eqref{2S}.   Following the time-evolution of the wave packet along the structure we determine the splitting ratio from the transmission 
into the outgoing upward-running channel, {\it i.e.} in positive y-direction, for rectangular intersections (see also Fig. \ref{texture}).

\subsection{Rectangular intersections}

Fig. \ref{wppwithV} (a) gives a series of snapshots showing the wave packet as it propagates horizontally along the in-channel and splits more or less symmetrically into two outgoing wave packets
traveling along the vertical channels.   For rectangular two-soliton magnetic textures of the form Eq. \eqref{2S} there are the domain width $\lambda$ and the angle $\phi$ characterizing the in-plane magnetization across the domain wall which can be varied. 
For the first  simulations of these dependencies for rectangular intersections shown in Figs. \ref{is1} and \ref{is2},  the domain wall width is chosen to be a relatively large $\lambda = 25$ nm compared to the perpendicular confinement of the wave packet $\approx100$ nm to bring out more clearly the influence of the details of the in-plane components of the magnetization on the wave packet propagation.
\begin{figure}[!t]
\includegraphics[width=8cm]{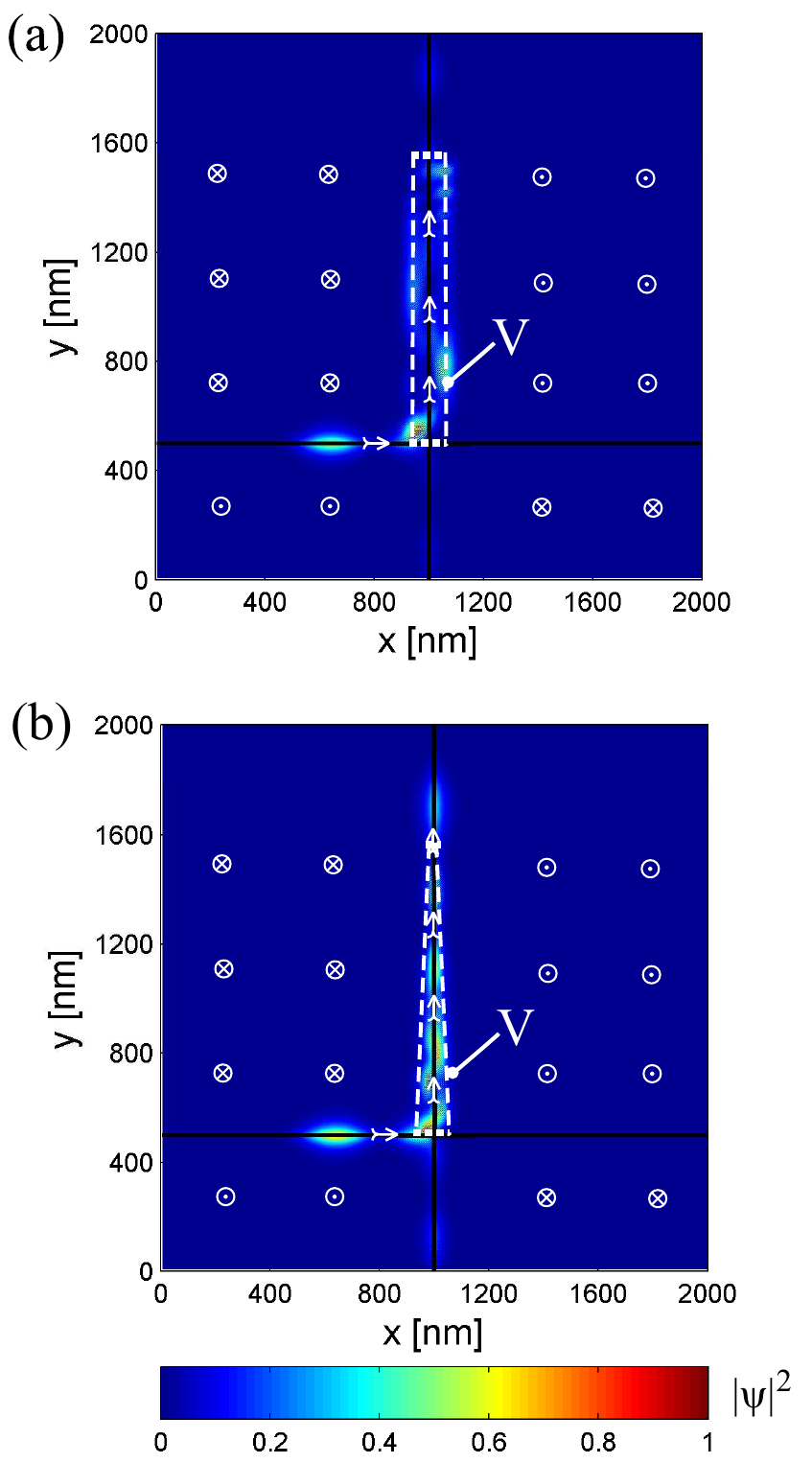}
\caption{(Color online). Snapshots of the wave packet propagation in a $90$-degree domain wall intersection: (a) with rectangular electrode, (b) with tailored electrode and a bias of $15$ mV .  The color (or brightness variation) shows the probability density $|\psi|^2$ (see color-bar). The black contour lines and the white crossed and dotted circles show the $M_z$-component of the magnetization.}\label{wppwithtriangleVn}
\end{figure} 

In Fig. \ref{is2} we show the effects on the transmission into the upper channel when the domain wall width $\lambda$ and the magnetization angle $\phi$ are varied.  Here, the wave packet is prepared with an energy mean value $E=0$. The color-coded figures show grid lines (in gray), altitude lines (black) as well as the altitude (as interpolated color map) which should allow the reader to assign a transmission value to every parameter combination shown in the plot. One observes that even for relatively large domain wall widths the influence of the in plane magnetization to the splitting behavior is moderate and, to a very high accuracy, a rectangular intersection  provides a 50-50 beam splitter for domain wall states over a wide energy range.   This is important for robustness of the splitting ratio under local imperfections.

\begin{figure} [t!]
\includegraphics[width=8.6cm]{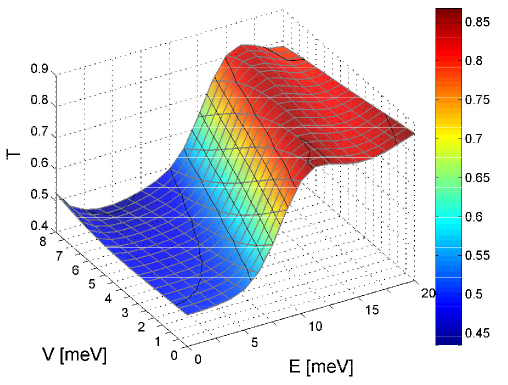}
\caption{(Color online). Relative transmission to the upper channel in a rectangular domain wall intersection as a function of the mean energy of the wave packet $E$ and the gating potential $V$ as shown in Fig. \ref{wppwithV} (b). The wall width is chosen to be $\lambda = 25$ nm and $\phi=0$.}\label{isV1n}
\end{figure}

This situation can be changed, when an electrode is placed 
asymmetrically onto the junction,  as sketched in Fig. \ref{wppwithV} (b).  
The chosen bias value  is $15$ mV. Note that a static potential cannot close the channel or revert the propagation direction. However, if sufficiently large it provides a mixing of 2D surface states with the channel state and, when spatially confined perpendicular to the out channel, provides a wave guide-like channel, as is shown in the simulation in Fig. \ref{wppwithV} (b).  This confinement effect, arising from wave number mismatch, has been discussed in the literature.\cite{yokoyama}  Here it is used, to control the splitting ratio which is shown for varying $V$ and $E$ and  $\lambda = 25$ nm and $\phi=0$ in Fig. \ref{isV1n}.  Dynamically, the confining effect by the (rectangular) gate resembles total internal reflection which arises under glancing incidence onto the potential wall. Tailoring the shape of the electrode helps in effective ``funneling" of fermions back into the domain-wall state.  This is demonstrated in Fig. \ref{wppwithtriangleVn} (a) and (b) in which we compare the effect of a rectangular versus a triangular bias region.   

Note that asymmetric  biasing of one out channel shows  influence on the splitting ratio only when the applied voltage is sufficiently high to energetically move the Dirac fermion out of the magnetization gap (see Figs. \ref{wppwithtriangleVn} and \ref{isV1n}). Otherwise it cannot influence significantly the propagation because the addition of a scalar potential does not change the group velocity due to the linear dispersion relation of the channel states.

\subsection{45-degree intersections}

\begin{figure} [!t]
\includegraphics[width=8cm]{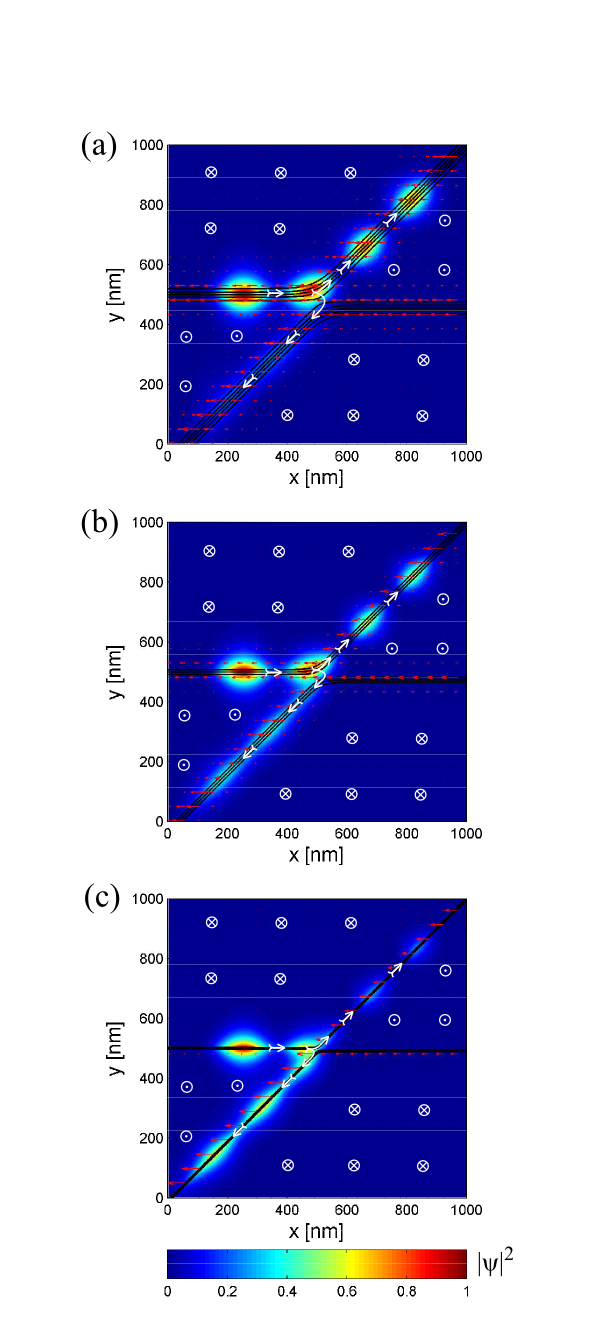}
\caption{(Color online). Snapshots of the wave packet propagation in a $45$-degree domain wall with a wall width of (a) $\lambda=25$ nm, (b) $\lambda=15$ nm and (c) $\lambda=5$ nm. $E=\phi=\pi$. The color (or brightness variation) shows the probability density $|\psi|^2$ (see color-bar). The black contour lines and the white crossed and dotted circles show the $M_z$-component of the magnetization. The vector plot shows the direction and magnitude of the in-plane magnetization $(M_x,M_y)$.}\label{wpp}
\end{figure}

\begin{figure} [!h]
\includegraphics[width=8.6cm]{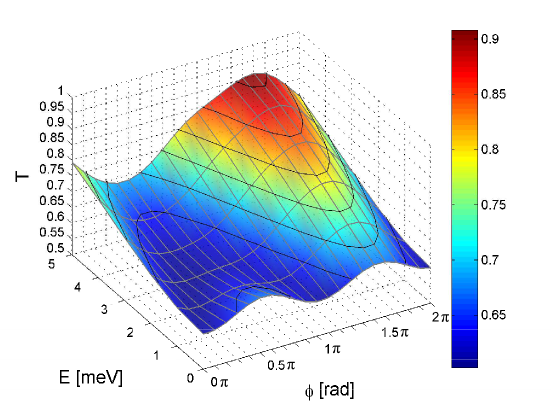}
\caption{(Color online). Relative transmission to the upper channel in a 45-degree domain wall intersection with wall width $\lambda = 25$ nm as a function of the in-plane magnetization angle $\phi$ and the energy mean value $E$ of the  wave packet.}\label{is3}
\end{figure}
\begin{figure} [!h]
\includegraphics[width=8.6cm]{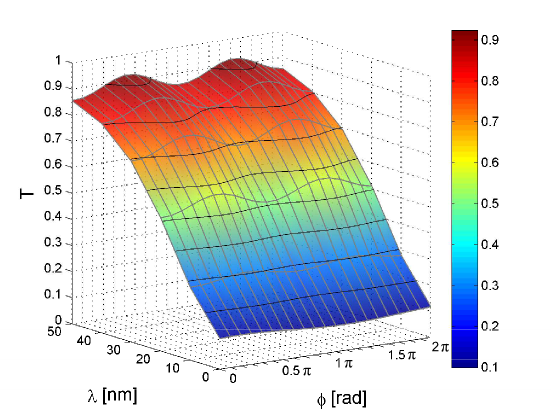}
\caption{(Color online). Relative transmission to the upper channel in a 45-degree domain wall intersection as a function of the in-plane magnetization angle $\phi$ and wall width $\lambda$. The in propagation-direction Gaussian shaped initial wave packet is prepared with an energy mean value of $E=0$.}\label{is4}
\end{figure}
\begin{figure} [!h]
\includegraphics[width=8.6cm]{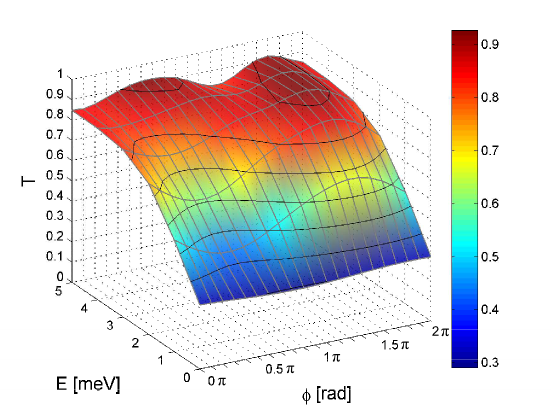}
\caption{(Color online). Relative transmission to the upper channel in a 45-degree domain wall intersection as a function of the in-plane magnetization angle $\phi$ and wall width $\lambda$. The in propagation-direction Gaussian shaped initial wave packet is prepared with an energy mean value of $E=5$ meV.}\label{is5}
\end{figure}
We now turn to  45-degree intersections and study them as a function of $\phi$ (domain-wall type), $\lambda$  (channel width), and $E$ (mean energy of the wave packet). 
Fig. \ref{wpp}  shows snapshots of the wave packet as it propagates along the in-channel and splits up at the intersection.  It is observed that the preferred out-channel depends on the channel width.  For wide channels the upper out-channel is preferred, while for narrow channel width, the  lower channel is the preferred exit.  This trend arises from an increased overlap of the in-channel wave-packet with the lower out-channel as the channel width decreases, whereas for wide channels, the path of lower momentum transfer wins out.   This effect therefore also displays an energy dependence, in contrast to the rectangular case.  

Results from a more systematic analysis are summarized in the following figures.  
In Fig. \ref{is3} we show the transmission as a function of the in plane magnetization angle $\phi$ and the energy mean value $E$ for $\lambda = 25$ nm.
In Fig. \ref{is4} we vary the wall width $\lambda$ and the in-plane magnetization angle $\phi$ using $E=0$. Fig. \ref{is5} shows the same setup but for $E=5$ meV. Compared to the rectangular intersections (Figs. \ref{is2} and \ref{is3}) the influence of the in-plane magnetization direction is more pronounced.  For a fixed interaction angle the most relevant parameter still is the wall width since it determines the degree of  asymmetry in the ``channel cross-talk" near the junction, as shown in quantitative detail in these figures.  As the fermion approaches the intersection it probes the surroundings and begins to leak into the lower out-channel before it detects the upper one.

\begin{figure} [!h]
\includegraphics[width=8.6cm]{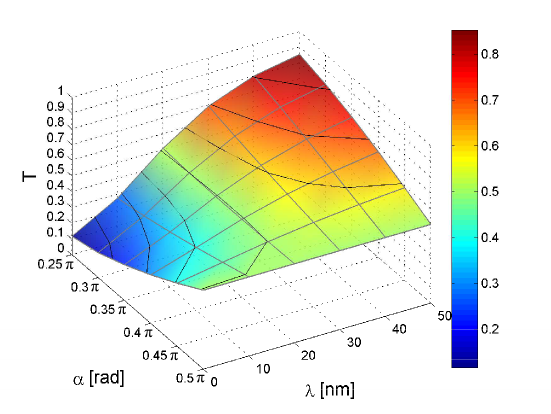}
\caption{(Color online). Relative transmission to the upper channel in a domain wall intersection where the intersection  angle $\alpha$ and the wall width $\lambda$ are varied. $E=0$ meV and $\phi=0$.}\label{is6}
\end{figure}
\begin{figure} [!h]
\includegraphics[width=8.6cm]{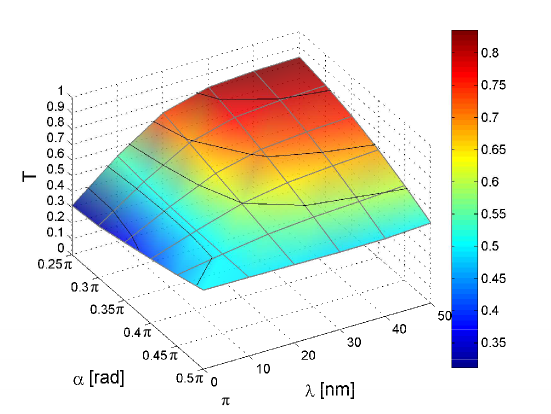}
\caption{(Color online). Relative transmission to the upper channel in a domain wall intersection where the intersection  angle $\alpha$ and the wall width $\lambda$ are varied. $E=5$ meV and $\phi=0$.}\label{is7}
\end{figure}
\begin{figure} [!h]
\includegraphics[width=8.6cm]{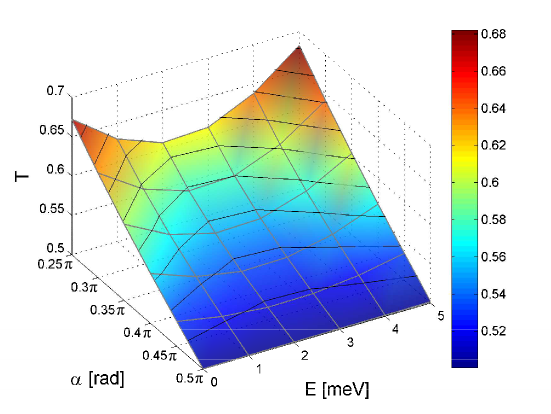}
\caption{(Color online). Relative transmission to the upper channel in a domain wall intersection where the intersection  angle $\alpha$ and the mean energy $E$ are varied.  $\phi=\pi/2$ and $\lambda=25$ nm.}\label{is9}
\end{figure}
\begin{figure} [!h]
\includegraphics[width=8.6cm]{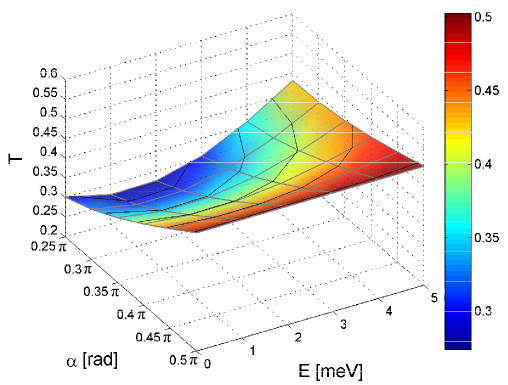}
\caption{(Color online). Relative transmission to the upper channel in a domain wall intersection where the intersection  angle $\alpha$ and the mean energy $E$ are varied.  $\phi=\pi/2$ and $\lambda=10$ nm.}\label{is10}
\end{figure}

\subsection{General angles of intersection}

Referring to Fig. \ref{texture}, general angles of intersection $\alpha$ are investigated.  In conjunction with the domain-wall thickness the angle of intersection plays the dominant role for establishing the splitting ratio of such an intersection.  
Results are summarized in the following figures.
In Fig. \ref{is6} we vary $\alpha$ and $\lambda$ for $E=\phi=0$. For Fig. \ref{is7} we change $E$ to $E=5$ meV. 
Fig. \ref{is9} shows the variation of $E$ and $\alpha$ where $\phi=\pi/2$ and $\lambda = 25$ nm. For $\lambda=10$ nm this setup shows a quite different behavior Fig \ref{is10}.
It is remarkable that for small
domain-wall widths, relative to the spatial extent of the Dirac fermion wave packet,  the transmission into the upper channel decreases for a more acute angle $\alpha$.   
For large widths, on the other hand,  the transmission into this out channel is favored, as one would suspect ``intuitively" (see Fig. 12 and 13).
This effect is most dominant for wave packets with a center energy of $E=0$.
The two competing effects responsible for this behavior are the difference in spatial (time-dependent) overlap of the incident wave-packet
with the two outgoing channels  and the ``wave-vector" mismatch at the intersection.  As already pointed out for the $\alpha=45^o$, the behavior of narrow channels is dominated by the former, while wide channels act ballistic, favoring low momentum transfer.  In other words, when going from narrow to wide channels, the Dirac fermion scattering behavior changes from quantum-mechanical wave-like to ``classical" particle-like.     Note that, as per Eq. \eqref{Y'},  the (stationary) transverse extent of the wave packet is determined by the asymptotic values of the magnetic texture, while the width of the domain wall independently is determined by the relative importance between anisotropy and exchange contribution in the free energy functional of the magnetization, Eq. \eqref{Fmag}.  Hence, in principle, structural design allows for 
both situations to occur.   

\section{Summary and conclusions}\label{SUC}

In summary, we have investigated numerically the dynamics of domain-wall Dirac fermions at the intersections of two linear ferromagnetic domain walls.  To model a realistic stable magnetic intersection texture we have used a  two-soliton solution of the sine-Gordon equation which establishes the optimality condition for a minimum of  a free energy functional accounting for exchange and anisotropy.   The time-dependent  analysis of chiral fermion propagation is based on a staggered grid numerical scheme to solve the effective (2+1)D Dirac equation. Developed for this particular purpose, it is constructed such that fermion doubling is avoided and absorbing boundary conditions in form of regions of imaginary scalar potential can be incorporated.  Details of this model and an extension to (3+1)D can be found elsewhere.\cite{Hammerup}  

The properties of such magnetic intersections as a one-way beam splitters  for chiral fermions have been investigated.   Based on this study, we 
can conclude that the splitting ratio for domain-wall fermions at the intersection depends strongly on the angle of intersection and, in case of non-rectangular intersections, on the width of the domain-wall. The latter determines the importance of the wave nature of the fermion onto the transmission behavior: quantum-tunneling dominates the behavior at the intersection when the (transverse) localization length of the domain-wall fermion is large compared to the channel width.   For wide domain walls the Dirac fermion behavior at the intersection becomes particle-like.   The type of domain wall, as well as the mean energy of the Dirac fermion wave packet, have a weaker influence on the splitting ratio.  The former is characterized  by the angle of the in-plane magnetization when its z-component goes through zero and allows one to compare Bloch- to N{\'e}el-type intersections.  Although experimental setups will have imperfections regarding the magnetic structure our results should give a qualitative guide for the splitting behavior in such structures, as long as defects do not destroy the channel states. 

External control of of the splitting ratio may become feasible in experiment via asymmetrically biased electric gates placed near the junction.  Using specially tapered electrodes, ``funneling" of laterally confined  fermions back into domain-wall states is shown in our simulation.   

 Here we have confined our analysis to the intersection of two linear domain walls.   If more complex magnetic  textures on topological insulators can be realized in experiment,  domain-walls can be used as chiral Dirac fermion waveguides, with their mutual intersections acting as beam-splitters as, for example, in a proposal for a domain-wall electric-gate controlled fermion interferometer. \cite{hammerAPL}

\begin{acknowledgments}
The work is supported by the Austrian Science Foundation under Project No. I395-N16.
\end{acknowledgments}

%
\end{document}